\documentclass[aps,pre,twocolumn,10pt]{revtex4-1}

\usepackage[dvips]{graphicx}
\usepackage{amssymb,amsfonts,amsmath}
\usepackage{color}
\usepackage{units}
\usepackage{colortbl}

\allowdisplaybreaks[1]
\newcommand{\comment}[1]{}

\begin{document}
\title{Analytical investigation of self-organized criticality in neural networks}

\author{Felix Droste}
\email[]{felix.droste@bccn-berlin.de}
\affiliation{Bernstein Center for Computational Neuroscience, Haus 2, Philippstrasse 13, 10115 Berlin, Germany}
\author{Anne-Ly Do}
\affiliation{Max Planck Institute for the Physics of Complex Systems, N\"othnitzer Str. 38, 01187 Dresden, Germany}
\author{Thilo Gross}
\affiliation{University of Bristol, Department of Engineering Mathematics, Merchant Venturers Building, Bristol BS8 1UB, UK}

\date{\today}

\begin{abstract}
Dynamical criticality has been shown to enhance information processing in dynamical systems, and there is evidence for self-organized criticality in neural networks. A plausible mechanism for such self-organization is activity dependent synaptic plasticity. Here, we model neurons as discrete-state nodes on an adaptive network following stochastic dynamics. At a threshold connectivity, this system undergoes a dynamical phase transition at which persistent activity sets in. In a low dimensional representation of the macroscopic dynamics, this corresponds to a transcritical bifurcation. We show analytically that adding activity dependent rewiring rules, inspired by homeostatic plasticity, leads to the emergence of an attractive steady state at criticality and present numerical evidence for the system's evolution to such a state.
\end{abstract}

\maketitle
Information processing systems often exhibit optimal computational capabilities when their parameters are tuned to \emph{critical states} associated with phase transitions \cite{Langton90, Kinouchi2006, Legenstein2007, Juelicher}. 
It therefore appears likely that our brains operate at criticality \cite{Beggs2008}.
Although still hotly debated in neuroscience, the hypothesis of neural criticality is supported by recent experiments. 
Power-law distributions indicative of critical behaviour were observed in slices of rat cortex \cite{Beggs2003, Beggs2004, Plenz2007} as well as EEG \cite{Novikov97, Freeman2000, Meisel2012}, fMRI \cite{Kitzbichler2009}, and EcoG \cite{Meisel2012} measurements in humans.     

In the light of the experimental corroboration of neural criticality it is interesting to ask how a biological system can robustly self-tune its parameters to a critical state.
A likely answer is found in the study of adaptive networks \cite{Gross2008,ANBook}, a class of models in which the dynamics on a network coevolves with the network structure. 
Already in 1998, it was noted that adaptive networks with slowly evolving topology can self-organize to a state where the dynamics on the network are critical \cite{Christensen98}. 
Adaptive self-organized criticality (aSOC) was subsequently demonstrated conclusively in a simple Boolean network model \cite{BornholdtRohlf} and then studied in detail in neural models \cite{Ebel2002, BornholdtRoehl, Levina2007, Burda2008, Levina2, Meisel09, MacArthur2010}.

We note that insights in aSOC may not only advance our understanding of natural neural networks, but may reveal an important design principle for electronic computers: 
Within the next ten years, current manufacturing processes will hit fundamental boundaries \cite{IEEERoadMapforSemiconductors}.  
Continued progress will require utilising nanoscale components (e.g.~nanotubes, nanowires, biomolecules) that cannot be positioned precisely with present-day techniques for manufacturing large-scale integrated systems.
It is thus conceivable that future computers may consist of active (nano-)elements that are deposited randomly and then left to self-tune to a critical state, where meaningful information processing is possible. 
A promising mechanism for such self-tuning is provided by aSOC.
Importantly, this mechanism does not require the rewiring of physical interconnections, but can be achieved already by local changes of conductivity between elements \cite{Meisel09} that have recently been demonstrated experimentally \cite{Jo10}.
      
Despite the prominent role of aSOC for information processing systems in both biology and technology, our understanding of the phenomenon is still limited.  
In many aSOC models, the critical state is identified by showing that certain quantities follow power-law distributions. 
However, power laws can appear due to other mechanisms besides criticality.
It is thus desirable to make the corresponding dynamical phase transition directly accessible to analytical investigation. 

The aim of the present paper is to develop a better understanding of aSOC by means of a simple conceptual model.
We start in Sec.~\ref{sec:background} with a brief review of the biological background. In Sec.~\ref{sec:model}, we introduce a simple neuron model.
In Sec.~\ref{sec:mca}, we use a moment closure approximation \cite{Gross06} for deriving a low-dimensional system of ordinary differential equations (ODEs) that capture the emergent dynamics.
By analysing the bifurcation structure of these ODEs, we show in Sec.~\ref{sec:phasetrans} that the model exhibits a non-equilibrium phase-transition. 
In Sec.~\ref{sec:socmecha}, we discuss the conditions under which the chosen topological update rules drive the system toward the phase transition. 
Finally, in Sec.~\ref{sec:socana} we show analytically and numerically that the model indeed exhibits aSOC.

\section{Biological Background}\label{sec:background}
In biology, transmission of information between neurons occurs via cell contacts known as synapses.
Up to the order of $10^5$ such connections (with different partners) can exist on a single neuron.  
Synapses allow a pre-synaptic neuron to depolarize a post-synaptic neuron, similar to polar devices that transmit current only in one direction. 
The topology of interconnections can thus be captured by a directed network, in which the nodes correspond to neurons and the directed links correspond to synapses. 

On a short timescale, neurons encode information in an electric potential across their cell membrane. 
In the absence of inputs a neuron approaches a \emph{resting state} with a characteristic membrane voltage. 
Depolarization of the membrane due to input from other neurons can lead to a \emph{firing state} in which active mechanisms are used to emit a strong voltage pulse, which in turn excites connected neurons.  
After firing, the neuron enters a \emph{refractory state} in which no excitation is possible before it finally returns to the resting state. 

On a longer timescale, the strength of synapses changes depending on the activity of the connected neurons.
Thus, from an engineering perspective the synapse is a memristive element \cite{Chua71}.
In biological terms the processes affecting synaptic strength are collectively known as synaptic plasticity. 
Here, we specifically consider \emph{homeostatic synaptic plasticity} (HSP) \cite{Turrigiano98, Turrigiano2004}, which decreases the strength of synapses if the activity of a neuron is high, and increases the strength if the activity is low.  

\section{Discrete Neural Model}
\label{sec:model}
In the present paper we consider a directed network of $N$ nodes. 
At any time a given node is in one of three different states: resting (inactive, I), firing (F), or refractory (R).
We start from a random network in which a majority of the nodes are in the inactive state, whereas a small fraction is in the firing state.
The node states are then evolved according to the following rules: Firing nodes become refractory at a rate $i$, refractory nodes become inactive at rate $r$, and, for every link pointing from a firing neuron to an inactive neuron, the target neuron is set to the firing state at rate $p$.

The network topology is evolved according to a rule modelling synaptic plasticity: 
A firing node looses an incoming link at the rate \(l\) whereas new links are established between nodes at the rate \(g\). 
The creation and deletion of links is reminiscent of the formation of synaptic contacts in the developing brain, but is used here as a discretized model of the continuous changes of synaptic weight in the adult brain. By formulating the model in terms of discrete linking and unlinking events we avoid additional complications caused by real-valued link dynamics.

The conceptual model described above is one of the simplest conceivable settings that allows studying the interplay between the spreading of excitation and homeostatic topological evolution.
Excitation dynamics are modelled by stochastic transitions with constant rates.
As a consequence, a node can be excited by a single active neighbour, whereas in real neural systems, multiple inputs are needed.
Moreover, the time a node spends in firing state is exponentially distributed, while real action potentials have a well defined length.
However, as known from epidemiological modelling, none of these simplifications impair the validity of the model predictions\cite{AndersonMay91}.

\section{Low-dimensional approximation}
\label{sec:mca}
Let us now derive a low-dimensional description for the time evolution of macroscopic quantities in the limit of infinite \(N\). 
To this end, we employ a \emph{moment closure approximation} (MCA)\cite{Bauch05, Gross06, Zschaler2012}.
We denote the densities of nodes in a certain state by \([F], [I], [R]\), respectively. 
Similarly, we denote the per-neuron density of links from a node in state X to a node in state Y by \([XY]\). 

Consider $[F]$, the density of nodes in firing state:
It increases when a firing neuron excites an inactive node, which happens at rate  at rate $p [FI]$, and decreases when a firing node becomes refractory, which happens at rate $i [F]$.
By such reasoning, we obtain equations for the time evolution of $[F]$, $[I]$ and $[R]$, which however, do not form a closed system as they depend on the link density $[FI]$. 

One possibility to close the system of equations is to approximate the link density by the densities of the nodes involved, usually by assuming $[FI] \approx k [F][I]$, where $k$ is the mean degree of the network.
Note that in such a \emph{mean-field approximation}, correlations between neighbouring nodes are neglected.

Alternatively, it is possible to treat the link densities themselves as dynamical variables and derive evolution equations for them. 
In general, these equations depend on triplet densities, which can then be approximated by link and node densities.
This approach is often referred to as the \emph{pair approximation}; it is described in detail in Appendix~\ref{sec:ap:mca}.

Using the pair approximation and assuming a Poissonian degree distribution, we obtain the following ODE description of the system:
{ \small
\begin{subequations}\label{eqn:ODE}
\begin{align}
 \dot{[F]} &= p [FI] - i [F] \label{eqn:dFdt}\\
 \dot{[R]} &= i [F] - r [R] \label{eqn:dRdt} \\
 \begin{split}
 \dot{[FF]} &= -2 i [FF] + p \left( \tfrac{1\comment{\kappa}}{2} \tfrac{[FI]^2}{[I]}+ [FI] \right) -l [FF]\\
            &\quad + g [F] [F] \label{eqn:FF} 
 \end{split} \\
 \begin{split}
  \dot{[FI]} &= -i [FI] + p \left( \tfrac{[II][FI]}{[I]}-  \tfrac{1\comment{\kappa}}{2} \tfrac{[FI]^2}{[I]}- [FI] \right) + r [FR] \\
             &\quad + g [F] [I] 
 \end{split} \\
 \dot{[FR]} &= i \left([FF] - [FR]\right) + p \tfrac{[IR][FI]}{[I]}- r [FR] + g [F] [R] \\
 \dot{[II]} &= - p  \left(1 + \kappa\right) \tfrac{[II][FI]}{[I]} + r \left([RI] + [IR] \right ) + g [I] [I]\\
 \dot{[IR]} &= i [IF] - p  \tfrac{[IR][FI]}{[I]}+ r \left( [RR] - [IR] \right) + g [I] [R] \\
 \begin{split}
  \dot{[RF]} &= i \left( [FF] - [RF] \right) + p \comment{\kappa} \tfrac{[RI][FI]}{[I]}- r [RF] - l [RF]\\ 
             &\quad + g [R] [F] 
 \end{split} \\
 \dot{[RI]} &= i [FI] - p \comment{1\comment{\kappa}} \tfrac{[RI][FI]}{[I]}+ r ([RR] - [RI]) + g [R] [I] \\
 \dot{[RR]} & = i \left( [FR] + [RF] \right) - 2 r [RR] + g [R] [R] \label{eqn:RR}\\
 \dot{k} &= g - l [F] \label{eqn:dkdt}
\end{align}
\end{subequations}}
where the last equation explicitly captures the change of the mean degree $k$ of the network. Writing the equation for $k$ saves us from writing the longer equation for IF, due to 
\begin{align}
 \begin{split}
  [IF] &= k - [FF] - [FI] - [FR] - [II] - [IR] \\&\quad - [RF] - [RI] - [RR] \ .
 \end{split}
\end{align}
Similarly $[I]$ follows from the conservation relation for nodes
\begin{equation}
   [I] = 1 - [F] - [R]\ .
\end{equation}

\section{Non-equilibrium phase transition}
\label{sec:phasetrans}
Before we address the dynamics of the full system, let us first focus on the case  \(l = g = 0\), where the network is static. In this case, the right hand side of the equation of motion for $k$ vanishes such that $k$ becomes a control parameter. Below, we will refer to the system without topological evolution as the \emph{static network model} and to the complete system as the \emph{adaptive network model}. 

The static network has a trivial steady state \({\rm \bf x}^0 = ([F]^0, \cdots, [RR]^0)\) in which all nodes are inactive, i.e \([I]^0=1, [II]^0 = k, [F]^0 = [R]^0 = [FF]^0 = \dots = [RR]^0 = 0\). Depending on parameters, it may moreover have a non-trivial, active steady state, in which \([I]<1 \).     
The stability of the trivial steady state is determined by the spectrum of the Jacobian matrix ${\rm \bf J|_{{\rm \bf x}^0}}\in \mathbb{R}^{10\times10}$, where $J_{ij}=\partial \dot{x_i}/\partial x_j$. The steady state is asymptotically stable if all eigenvalues  of \({\rm \bf J}|_{{\rm \bf x}^0}\) have a negative real part \cite{Kuznetsov98}. 

If the variables $x_i$ are ordered as in Eqns.~\eqref{eqn:ODE}, the non-vanishing entries of \({\rm \bf J}|_{{\rm \bf x}^0}\) are
{ \small
\begin{align}
    J_{1,1} &= J_{7,3} = J_{7,4} = J_{7,5} = J_{7,6}  
     = J_{7,8} = J_{7,9} = -i\ ,\nonumber\\
 J_{1,4} &= J_{3,4} = p \ , \nonumber\\
  J_{2,1} &= J_{5,3} = J_{8,3}  = J_{9,4} = J_{10,5} = J_{10,8} = i\ , \nonumber\\
 J_{2,2} &=  J_{4,5} = J_{6,7} = J_{6,9} = J_{9,10} = r\ , \nonumber\\
 J_{3,3} &= -2 i\ , \nonumber\\
 J_{4,4} &= -i + (k - 1) p\ , \nonumber\\
 J_{5,5} &= J_{7,7} = J_{8,8} = -i - r\ , \nonumber\\
 J_{6,4} &= - 2 k p\ , \nonumber\\ 
  J_{7,10} &= - i + r\ , \nonumber\\
 J_{9,9} &= -r\ , \nonumber\\
 J_{10,10} &= -2r \ . \nonumber
\end{align}
}

In the following, we assume $i > 0$, $r > 0$ and $p > 0$.
The characteristic polynomial can then be factored into 7 linear factors with negative real roots and a remaining third order polynomial
\begin{equation}
 P(\lambda) = (-r - \lambda)^3 (-i - \lambda)^2 (-i-r - \lambda) (-2 r - \lambda) f(\lambda)\ ,
\end{equation}
where
\begin{equation}
\begin{split}
f(\lambda) &= i \left( i^2 + 2 i \left((1 - k) p + r\right) + (1 - 2k) p r \right)  \\
 &\quad + \left(5 i^2 + (1 - k) p  r + 3 i \left((1 - k p) + r\right) \right) \lambda \\
  &\quad + \left(4 i + (1 - k) p + r\right) \lambda^2 + \lambda^3\ . 
\end{split}
\end{equation}

In order to assess the stability of \({\rm \bf x}^0\) without explicitly calculating the roots of \(f(\lambda)\), we use the Routh-Hurwitz theorem \cite{RouthHurwitz}. It states that the roots of a polynomial \(p(x) = x^n + b_1 x^{n-1} + \dots + b_{n-1} x + b_n\) all have negative real parts if the \emph{Hurwitz determinants}
\begin{equation}
 \Delta_k = 
\begin{vmatrix}
b_1      &        1 &       0  &        0 & \dots  & 0\\
b_3      &      b_2 &      b_1 &        1 & \dots  & 0\\
\vdots   & \vdots   &   \vdots &   \vdots & \ddots & \vdots \\
b_{2k-1} & b_{2k-2} & b_{2k-3} & b_{2k-4} & \dots & b_k \\
\end{vmatrix},
\end{equation}
with \(k = 1 \dots n\), are all positive. Calculating the Hurwitz determinants of the Jacobian matrix \({\rm \bf J}|_{{\rm \bf x}^0}\) and evaluating the positivity conditions reveals that the trivial steady state is stable if and only if
\begin{equation}\label{criticalK}
 k < \frac{i}{p} + \frac{i + r/2}{i + r} =: k_{\rm c}. 
\end{equation}

At $k=k_{\rm c}$ the trivial steady state becomes unstable as the systems undergoes a transcritical bifurcation. In this bifurcation a nontrivial steady state enters the positive cone of the state space, becoming a physical solution. The transcritical bifurcation thus marks a transition between the trivial inactive state and an active state in which ongoing activity is observed.
This transition is in the same universality class as directed percolation.

\begin{figure}[t!]
 \includegraphics[width=\columnwidth]{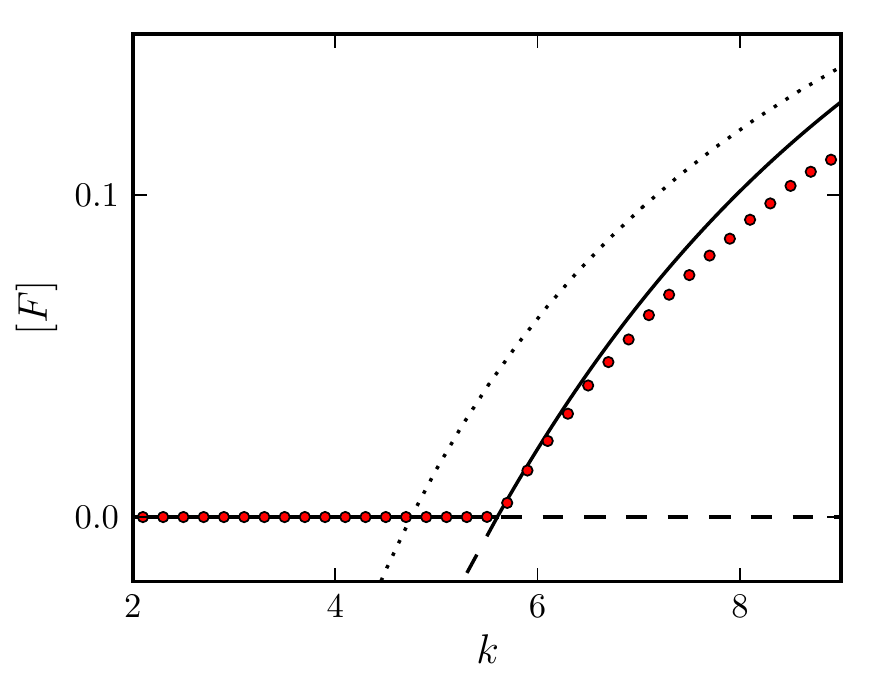}
 \caption{Bifurcation diagram for the static network. Plotted is the steady state density of firing neurons \([F]\) over the network's mean degree \(k\). 
 The solid line marks stable steady states of the static system, the dashed line unstable ones. At \(k = k_{\rm c} \approx 5.6\), the inactive steady state loses stability in a transcritical bifurcation. 
The respective transition from an inactive to an active phase is already observed in individual-based simulations with \(N = 10^6\) neurons (circles). Note that the critical \(k\) is nicely predicted by the link-level approximation Eqns.~\eqref{eqn:ODE}, but underestimated by a MCA at mean-field level (dotted lines). Parameters used were \(p = 0.2, i = 0.95, r = 0.4\). Nontrivial steady states were calculated using \texttt{AUTO}\cite{Doedel07}.}
 \label{fig:bifdiag}
\end{figure}

The role of the transcritical bifurcation is illustrated in a representative bifurcation diagram in Fig.~\ref{fig:bifdiag}. The diagram shows that the analytically predicted bifurcation point is in very good agreement with numerical results from agent-based simulations of the system. Further, numerical continuation of the ODEs~\eqref{eqn:ODE} shows good agreement close to the bifurcation point. By contrast, closing the MCA at mean-field level yields $[F]^0_{\rm mf} = r(1- i/(p k))/(i+r)$ for the non-trivial steady state. This can be seen to be a much poorer approximation and underestimates the bifurcation point.

\section{Local information and time-scale separation}
\label{sec:socmecha}

The previous section showed that the static network model exhibits a transcritical bifurcation. In the language of statistical physics this bifurcation constitutes a phase transition. To establish that the adaptive network model shows self-organized criticality we have to show that the evolution of the connectivity drives the system to the critical point.

In the discussion leading up to Fig.~\ref{fig:bifdiag} we treated $k$ as a parameter of the system. For studying the evolution of the connectivity we now consider $k$ as a dynamical variable that evolves according to Eq.~\eqref{eqn:dkdt}. By introducing dynamics in $k$ we change the dynamical system, which can potentially lead to a changed bifurcation diagram. Indeed, in general a diagram analogous to Fig.~\ref{fig:bifdiag} cannot be drawn for the adaptive network model because $k$ is not available as a parameter axis anymore. 

The pitfall described above is inherent in the concept of SOC: Criticality can only be cleanly defined in a different system, where the self-organization is absent. 
To circumvent this pitfall one has to demand that the self-organization acts on a slower timescale. In the example studied here this implies,
\begin{equation}\label{tss}
 g, l \ll p, i, r \ .
\end{equation} 
In this case, the bifurcation diagram of the static network model reappears as the bifurcation diagram of the fast-subsystem of the adaptive network model, where $k$ can again be treated as a parameter.   

The genesis of aSOC in our model requires additionally a second time scale separation. To see this let us revisit a widely held plausibility argument for aSOC\cite{BornholdtRoehl, Gross2008}. It is argued that in adaptive networks robust self-organized criticality is possible because the dynamics on the networks make information on the global topology available in every node. Based on this information the nodes can then infer the global phase and adjust the local topology accordingly. Indeed, in previous publications\cite{BornholdtRohlf, BornholdtRoehl} the networks nodes extracted the global phase information by integrating their dynamics over a long time. While some biological processes are known to enable such temporal averaging \cite{Desai99, Losonczy2008}, we explore an alternative explanation.

In our model, the local accessibility of the global phase information is limited, as both the dynamics on the static network as well as the topological updates are memoryless processes that depend only on the current state of a single link or node. 
Hence, neurons in the resting state do not possess any information about the global phase, as they occur in both phases.
By contrast, neurons in the firing state can infer the global phase from their local state, as their occurrence is restricted to the active phase.
To achieve aSOC despite the limited local accessibility of the global phase, links have to be created \emph{tentatively} as long as no definitive information is known, but destroyed \emph{decisively} once information about the global phase is available. 
This corresponds to a separation of timescale between the link creation and link deletion process
\begin{equation}\label{tss1}
 \epsilon := \frac{g}{l} \ll 1 \ .
\end{equation}
Note, that the twofold separation of time scales constituted by Eq.~\eqref{tss} and Eq.~\eqref{tss1} is typical for SOC models \cite{BornholdtRoehl,Levina2007,Meisel09}, although it is sometimes hidden in a non-local model definition \cite{Vespignani98, PathsToSoc}.
Our analysis reveals that it is indeed a necessary ingredient to achieve criticality without centralized control.

\section{Self-organized criticality}
\label{sec:socana}

Let us now show that the adaptive network model has a stable steady state, which in the double limit \(l\rightarrow0\), \(\epsilon \rightarrow 0\) coincides with the bifurcation of the static network model. 
We start by defining \({\rm \bf x}^* = ([F]^*, \cdots, [RR]^*, k^*)\) as a steady state of the adaptive network model~\eqref{eqn:ODE}. 
Evaluating the stationarity conditions \(\dot{k}|_{{\rm \bf x}^*} = 0, \dot{[F]}|_{{\rm \bf x}^*} = 0\) and \(\dot{[R]}|_{{\rm \bf x}^*} = 0\), we can immediately read off the steady state values 
\begin{equation}\label{stst}
	[F]^* = \epsilon, \quad [R]^* =   \epsilon i/r, \quad [FI]^* = \epsilon i/p \ .
\end{equation}
Therewith, the remaining equations (\(\dot{[FF]}|_{{\rm \bf x}^*} = 0, \cdots, \dot{[RR]}|_{{\rm \bf x}^*} = 0\)) become linear and can be solved by a computer algebra system. 
Due to the time-scale separation, higher-order terms in \(\epsilon\) and \(l\) can be dropped, and we obtain
{\small
\begin{subequations}\label{eq:staedystate}
\begin{align}
 [FF]^* &= \frac{1}{2} \epsilon \label{eqn:FFstar} \\
 [FR]^* &= [RF]^* = \frac{1}{2} \frac{i}{i + r} \epsilon \\
 [II]^* &= k_{\rm c} + \frac{r}{4 i (i + r)} l - \frac{i}{i + r} \left( k_{\rm c} + \frac{1}{2}  \right) \epsilon \\
 [IR]^* &= \frac{i}{r} \left( k_{\rm c} + \frac{1}{2}  \right) \epsilon \\
 [RI]^* &= \frac{i}{r} \left( k_{\rm c} - \frac{1}{2}  \right) \epsilon \\
 [RR]^* &= \frac{1}{2} \frac{i}{r} \frac{i}{i + r} \epsilon \label{eqn:RRstar} \\
\begin{split}
 k^* &= k_{\rm c} + \frac{r}{4 i (i + r)} l \\ &\quad + \left( \frac{i + r}{r} \left(\frac{1}{2} + 2 k_{\rm c}\right) -  \frac{i}{i + r} \left(1 + k_{\rm c} \right) \right) \epsilon  \label{eqn:kstar}\ .
\end{split}
\end{align}
\end{subequations}
}

For \(\epsilon, l \ll 1\), the steady state \({\rm \bf x}^*\) is always stable, which can be shown by calculating the characteristic polynomial of \({\rm \bf J}|_{{\rm \bf x}^*}\) and then checking the Hurwitz determinants for positivity (again, dropping higher-order terms in \(\epsilon\) and \(l\)). 
Moreover, the Eqns.~\eqref{eq:staedystate} reveal that in the double limit \(l \rightarrow 0, \epsilon \rightarrow 0\), \([II]^* \rightarrow k_{\rm c}\), while \([F]^*, \ldots,[RR]^* \rightarrow 0\), i.e., \({\rm \bf x}^*\) converges towards the transcritical bifurcation of the static network model. 

Recall that the ODE system describes the model system in the limit \(N \rightarrow \infty\); as observed recently, this is also the limit in which criticality may be expected for non-conserving dynamics \cite{Bonachela2009, Bonachela2010}. 
We thus conclude that for \(N \rightarrow \infty\), and \(l \rightarrow 0, \epsilon \rightarrow 0\), the HSP-inspired update rule self-organizes the adaptive system to criticality. \\

\begin{figure}[t!]
 \includegraphics[width=\columnwidth]{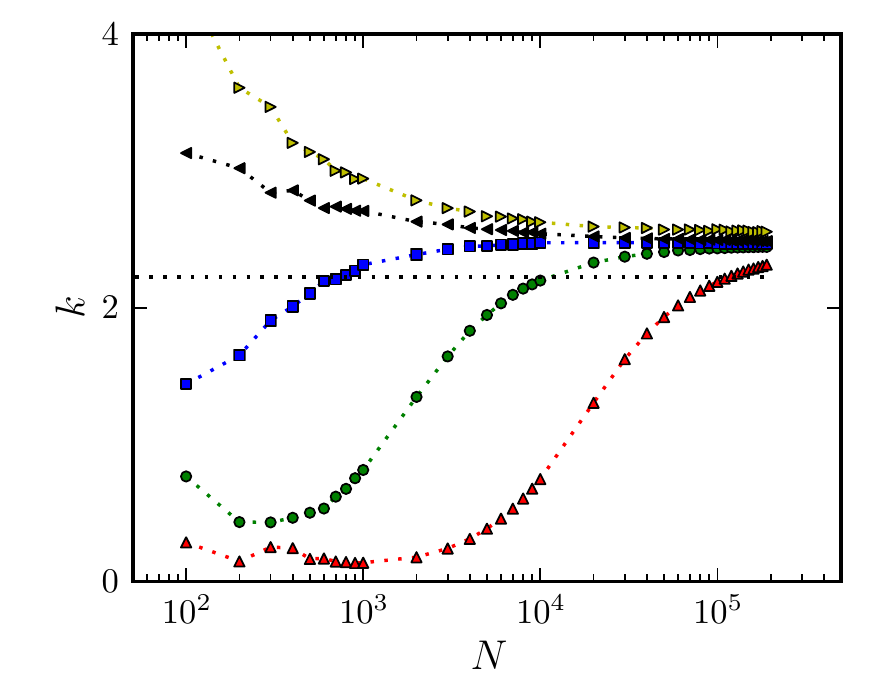}
 \caption{(Colour online) Mean degree \(k\) of evolved networks for different network sizes \(N\) and different spontaneous firing rates \(s = 1/(1000N)\) (yellow triangles pointing right), \(s = 1/(100N)\) (black triangles pointing left),  \(s = 1/(10N)\) (blue squares), \(s = 1/N\) (green circles), \(s = 10/N\) (red triangles pointing up). The dotted line marks the analytical steady state value for \(l = 0.01, \epsilon = 10^{-4}\). Simulations were run for $10^7$ time units, plotted connectivities are averages over the last $5\cdot10^6$ time units.  Other parameters: \(p = 0.7, i = 0.95, r = 0.4\).}
 \label{fig:kofN}
\end{figure}

\begin{figure}[t!]
 \includegraphics[width=\columnwidth]{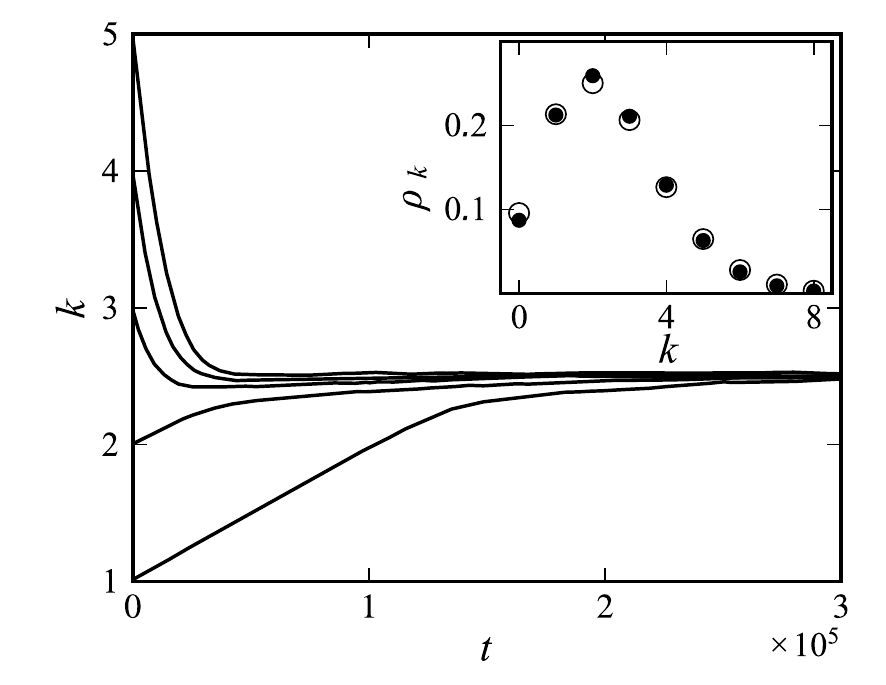}
 \caption{Time evolution of the mean degree of networks (\(N=10000, l = 10^{-3}, \epsilon = 0.01, s = 10^{-4}\)), starting from different initial connectivities. Inset: In-degree distribution of an evolved network after $5\cdot10^6$ time steps (\(N=10^5, l = 10^{-3}, \epsilon = 10^{-3}, s = 10^{-5}\), open circles) and Poissonian distribution around the same mean (filled circles). Other parameters: \(p=0.7, i=0.95, r=0.4\) in both cases.}
 \label{fig:ts}
\end{figure}

\begin{figure}[t!]
 \includegraphics[width=\columnwidth]{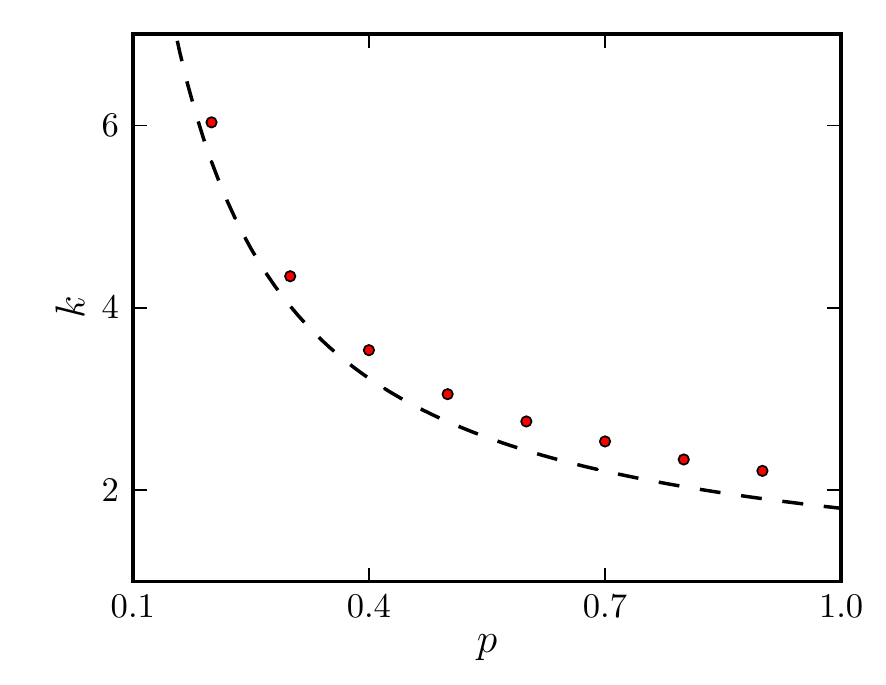}
 \caption{Mean degree \(k\) of evolved networks for different values of \(p\). The dashed line marks the analytically obtained critical connectivity \(k_{\rm c}\). Simulations were run for $10^8$ time units. Parameters: \(N = 10^4, l = 10^{-6}, \epsilon = 0.0015, s = 10^{-7}, i = 0.95, r = 0.4\)}
 \label{fig:kp}
\end{figure}

\begin{figure}[t!]
 \includegraphics[width=\columnwidth]{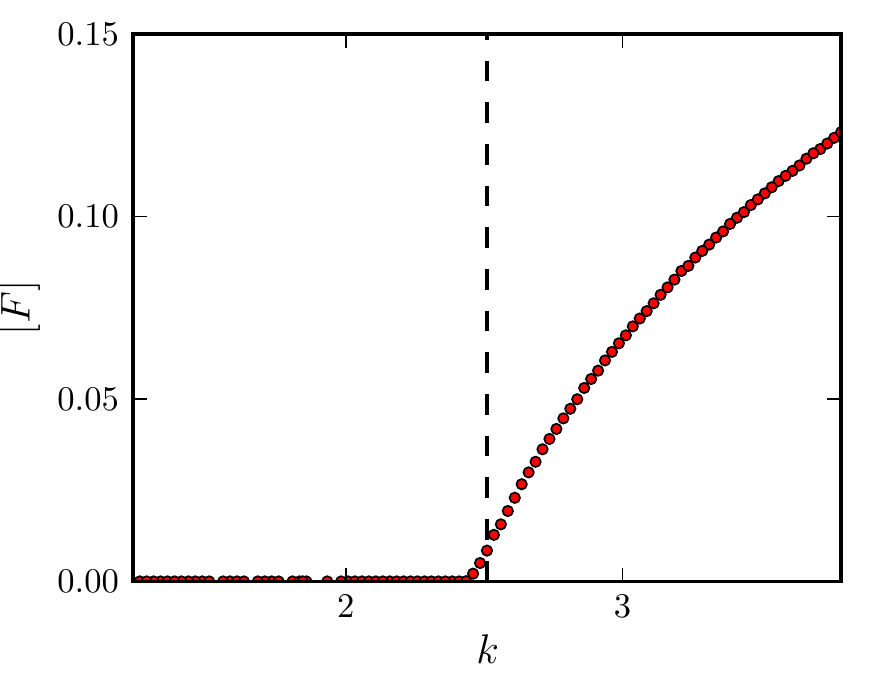}
 \caption{Average density of firing neurons (after transients) around the evolved connectivity (marked by vertical dashed line). Values above or below the evolved connectivity were obtained by adding (removing) random links to (from) the evolved topology. The network was evolved over $5\cdot10^6$ time units. Parameters: \(N = 10^5, p = 0.7, i = 0.95, r = 0.4, l = 0.01, \epsilon = 10^{-4}, s = 1/(100 N)\). }
 \label{fig:scan_super}
\end{figure}

To confirm aSOC in the discussed model, we ran individual-based simulations. 
We start with a network of \(N\) nodes, each of which has an outgoing connection to any other node with probability \(k_0/N\). We initialize 5\% of all nodes in the firing state, all others in the inactive state. Then, nodes and links are evolved according to the rules described in Sec.~\ref{sec:model} using the Gillespie algorithm\cite{Gillespie77}.

For finite $N$, the simulations tend toward the absorbing inactive state due to demographic stochasticity.
To compensate for the finite size effect, we include an additional process: Inactive nodes fire spontaneously at rate \(s\).
This process has an immediate biological interpretation as it reminds of the spontaneous activity observed in neurons. 
To reconcile the simulations with the low-dimensional description~\eqref{eqn:ODE}, \(s\) has to vanish in the \(N \rightarrow \infty\) limit. 
A plausible assumption is \(s = c/N\), where \(c\) can be chosen arbitrarily. 
We have verified that for sufficiently large system sizes the particular choice of \(c\) does not significantly influence the evolving connectivity (cf.~Fig.~\ref{fig:kofN}).

We start in Fig.~\ref{fig:ts} by plotting the time evolution of the mean degree \(k\) in networks with different initial configurations. 
All networks approach the same connectivity, which corroborates that the adaptive network model has exactly one stable steady state to which it converges irrespective of initial conditions. 

The inset of Fig.~\ref{fig:ts} shows that the in-degree distribution stays Poissonian throughout topological evolution and thus retrospectively corroborates the assumption made in the derivation of the MCA. 

So far we have shown that the network self-organizes to a unique value of the connectivity. It remains to test whether this value is indeed the critical connectivity.
As a first test, we compare the connectivity of the networks evolved in numerical simulations with the critical connectivity \(k_{\rm c}\) predicted by the analytical approximation from Eq.~\eqref{criticalK}. 
As shown in Fig.~\ref{fig:kp} the numerical result agrees with the analytical estimate of the critical state except for a small but systematic deviation. This discrepancy can be understood considering Eq.~\eqref{eqn:kstar}. The second and third term of this equation are positive for all $i,r,k_{\rm c},l,\epsilon>0$. Hence, the systematic deviation $k^* >k_{\rm c}$ is consequence of the networks' topologies being evolved with small but finite rates $\epsilon, l$.

As a second test, we directly probe the dynamics on the evolved networks. 
For this purpose, we first evolve the topology until the mean degree has reached a stationary state. Then, we deactivate the adaptive addition and  deletion of links but manually add (delete) links from the network. After every addition (deletion) we let the dynamics on the network reach a stationary level and record the average number of active neurons. 
As shown in Fig.~\ref{fig:scan_super}, this procedure recreates the phase diagram of the system numerically. It thereby provides direct evidence of the criticality of the evolved state.  
The slight displacement from the critical point can be understood recalling the analytical results given above: 
According to Eq.~\eqref{stst}, $[F]^*=\epsilon$ in the evolved network. 
Hence, the displacement toward the active regime can be attributed to the small but finite rates \(\epsilon, l\) used in the simulations.

\section{Discussion}
In summary, we have shown that activity-dependent synaptic plasticity self-organizes a neural network to criticality, provided that driving is slow and there is an appropriate separation of timescales between potentiation and depression. 

We have considered a simple discrete neural network model and used moment-closure approximation to derive a low-dimensional ODE representation, in which the dynamical phase transition becomes manifest as a transcritical bifurcation. 
By adding activity dependent rewiring rules we obtained an adaptive network model that has one attractive steady state. 
We have shown that, in the limit of infinitesimally slow topological adaptation, this steady state coincides with the bifurcation of the static network model, and, thus, that the adaptive model displays aSOC. 

We emphasize that the particular choice of the parameters $p$, $r$, and $i$ does not affect the qualitative model behaviour but only the quantitative predictions, in particular the evolved connectivity:
The probability that a firing node excites a specific neighbouring node is $p/i$. 
It is apparent from Eq.~\ref{criticalK} that choosing higher values for $p/i$ will result in lower evolved connectivities $k_c$.
In order to keep simulations computationally feasible, which specifically means maintaining a low number of links, we have chosen unphysiologically high values for $p/i$.
It can be seen from our analytical treatment that this affects neither the existence of a steady state at criticality nor the stability of this state.

Our work provides a conceptual angle on self-organized criticality of neurally inspired models. Although the phenomenon has been demonstrated in several earlier works \cite{Levina2007, MeiselThesis}, the simplified model studied here is the first system in which it is analytically tractable. Using dynamical systems and bifurcation theory, aSOC can be established more rigorously than in more realistic models and experiments, where evidence for criticality is mainly provided by the numerical observation of power laws. We believe that this approach will prove useful in the context of other, more complex dynamical phase transitions. For example, it is conceivable that a similar strategy may help to elucidate the recently observed self-organization to the onset of synchronous activity\cite{Meisel09}. We hope that the approach can thus contribute to settling an ongoing debate regarding the existence and function of criticality in the brain \cite{Beggs2008} and potentially to the development of self-organising electronic circuits. 


\begin{appendix}
\section{Moment-closure approximation}
   \label{sec:ap:mca}

As described in Sec.~\ref{sec:mca}, the time evolution of node densities is given by
\begin{subequations}
\begin{align}
 \dot{[F]} &= p [FI] - i [F],\\
 \dot{[R]} &= i [F] - r [R]. 
\end{align}
\end{subequations}

For an expansion beyond mean-field level, we need to derive equations for the time evolution of link densities.
There are four ways a link of a certain type be created or destroyed: 1) One of the link's nodes changes independently of others, 2) the link is added or removed as a result of topological evolution 3) one of the nodes is excited by the other via the link in question, or 4) one of the nodes is excited via a \emph{different} link. Processes of type 1), 2) and 3) can be understood on the level of nodes and links. To understand 4), consider that activity in a given focal link, say, excitation along an FI link, does not only affect the links itself, but also links connecting to it, here for instance other links connected to the I-node. Thus, processes of type 4) can only be captured if the density of subgraphs with two links, so-called \emph{triplets} are taken into account.

Consider the change in the density of FI links,
\begin{align}
 \begin{split}
  \dot{[FI]} &= p \left( [F{>}I{>}I] - [FI] - [F{>}I{<}F] \right) \\
             &\quad - i [FI] + r [FR] + g[F][I],
 \end{split}
\end{align}
where \([X{>}Y{>}Z]\) denotes the density of directed triplets, with the directionality of links indicated by ``\({<}\)''and ``\({>}\)'', respectively.
The three rightmost terms are of type 1) and 2). 
The others can be understood as follows: a \(\rm \underline{F{>}I}{>}I\) triplet turns into a \(\rm \underline{F{>}F}{>}I\) triplet at rate $p$, increasing the number of FI links by one (the underline denotes the link on which the update rule is applied). At the same rate, the F in \(\rm \underline{FI}\) links excites the I, which decreases the number of FI links by one. Finally, \(\rm \underline{F{>}I}{<}F\) triplets change to  \(\rm \underline{F{>}F}{<}F\) at rate \(p\), destroying the FI link \emph{not} underlined. In the last transition, there are actually \emph{two} FI links being destroyed, but we have already counted one of them as a process of type 3). Evolution equations for other types of links are obtained using the same reasoning.

The pair approximation closes the system of equations by approximating the occurring triplets densities in terns of link densities.
Let us start with a triplet of the type \(\rm X{>}Y{>}Z\). 
It consists of one XY link, which we know occurs with densities $[XY]$. If we assume that the states of next-nearest neighbours are uncorrelated, we can calculate the expected number of links from the Y node to a Z node as $[YZ]/[Y]$.  
We can thus approximate the triplet density as
\begin{equation}
[X{>}Y{>}Z] \approx \frac{[XY][YZ]}{[Y]}.
\end{equation}

Triplets of the type \(\rm X{>}Y{<}Z\) are approximated similarly: 
Again, the triplet contains an XY link, which occurs with density $[XY]$.
To calculate the expected number of ZY links connected to the central Y node, we now need to take into account that we already know that one incoming link is a XY.   
Thus, we need to consider the \emph{mean excess degree} $q$, i.e. the expected number of incoming links that the Y has \emph{in addition} to the XY link.  
Assuming that each of these links is a ZY link with probability $[ZY]/(k [Y])$, we can write
\begin{equation}
\label{eq_mc_mitkappa}
 [X{>}Y{<}Z] \approx \frac{q}{k} \frac{[XY][YZ]}{[Y]}. 
\end{equation}
Note that for symmetric triplets (\(\rm X{>}Y{<}X\)), Eq. (\ref{eq_mc_mitkappa}) has to be modified to avoid double-counting:
\begin{equation}
 [X{>}Y{<}X] \approx \frac{q}{2k} \frac{[XY]^2}{[Y]}.
\end{equation}

Assuming that the in-degree distribution of the evolving network is Poissonian, we can simplify the triplet approximations further. For Poissonian degree distributions, $q=k$ \citep{Newman01}, which leads to the expressions in Eq.~\eqref{eqn:ODE}. Note that the assumption is confirmed by the numerical results in Sec.~\ref{sec:socana}.

\end{appendix}

\end{document}